\documentclass{article}
\usepackage{xr}
\usepackage[utf8]{inputenc}
\usepackage{graphicx}
\usepackage{amsmath}
\usepackage{makecell}
\usepackage{algorithm2e}
\usepackage{xcolor}
\usepackage{tikz} 
\usetikzlibrary{positioning,arrows.meta,quotes,bayesnet,matrix}
\usepackage{xcolor}
\usepackage{caption}
\usepackage{subcaption}
\usepackage{capt-of}
\usepackage{bbm}
\usepackage{amssymb}
\usepackage{stackrel}
\usepackage{pifont}
\usepackage[numbers]{natbib}
\newcommand{\paren}[1]{\mathopen{}\left({#1}_{{}_{}}\,\negthickspace\right)\mathclose{}}
\newcommand{\bracket}[1]{\mathopen{}\left[ {#1}_{{}_{}}\,\negthickspace\right]\mathclose{}}

\newcommand{\Var}{\mathbbm{V}\textrm{ar}}
\newcommand{\E}{\mathbbm{E}}

\usepackage{arxiv}

\usepackage[utf8]{inputenc} % allow utf-8 input
\usepackage[T1]{fontenc}    % use 8-bit T1 fonts
\usepackage{hyperref}       % hyperlinks
\usepackage{url}            % simple URL typesetting
\usepackage{booktabs}       % professional-quality tables
\usepackage{amsfonts}       % blackboard math symbols
\usepackage{nicefrac}       % compact symbols for 1/2, etc.
\usepackage{microtype}      % microtypography
\usepackage{lipsum}		% Can be removed after putting your text content
\usepackage{graphicx}
\usepackage{natbib}
\usepackage{doi}
\author{
  Elouan Argouarc'h, Fran\c{c}ois Desbouvries, \\
  SAMOVAR \\
  Télécom SudParis, Institut Polytechnique de Paris \\
   91120 Palaiseau France\\
  \texttt{\{elouan.argouarch, francois.desbouvries\}@telecom-sudparis.eu} \\
}
\begin{document}

\title{Binary classification based Monte Carlo simulation}
\maketitle

\begin{abstract}
Acceptance-rejection (AR), 
Independent Metropolis Hastings (IMH)
or importance sampling (IS) Monte Carlo (MC) simulation algorithms 
all involve computing {\sl ratios} of 
probability density functions (pdfs). 
On the other hand, 
classifiers discriminate labeled samples 
produced by a mixture of two distributions
and can be used for approximating the {\sl ratio} of the two corresponding pdfs.
This bridge between simulation
and classification 
enables us 
to propose pdf-free versions of pdf-ratio-based simulation algorithms, 
where the ratio is replaced by a surrogate function computed via a classifier. From a probabilistic modeling perspective,
our procedure involves a structured energy based model
which can easily be trained and is compatible with the classical samplers.
\end{abstract}
\section{Introduction}
\label{intro}

If $a$ and $b$ are two positive numbers,
\begin{eqnarray}\label{fondamental}
r = \frac{a}{a+b} \in (0,1) \Leftrightarrow \frac{r}{1-r} = \frac{a}{b} > 0.
\end{eqnarray}
This identity has interesting consequences in Bayesian classification, machine learning and stochastic simulation. Indeed, if $a$ and $b$ are probabilities of two classes in a binary mixture context for a given sample, then ratio $\frac{a}{a+b}$ is the the posterior probability which provides with the class probabilities for a given sample, and can be approximated by a parametric classifier 
$r_\phi$ 
trained to distinguish between the two probability distributions. On the other hand, 
positive ratios $\frac{a}{b}$ play a key role in AR, IMH or IS techniques. Equation \eqref{fondamental} 
relates $r$ to such positive ratios,
and tells us that {\sl ratio $\frac{a}{b}$ can be computed exactly from $r$,
or, in practice, 
approximately from $r_\phi$,
without necessarily knowing $a$ nor $b$}.
This observation enables us 
to propose approximate versions 
of these samplers 
which rely on weaker hypotheses.

Let $\lambda, 1\!-\!\lambda \in (0,1)$ be the prior probabilities of two categories $k=1,0$,
distributed resp. 
$\sim$
$p_1$ and $p_0$.
Binary classification tries to distinguish samples 
from  mixture $\lambda p_1 + (1-\lambda)p_0$ 
by identifying the pdf which 
generated them. 
The appropriate way to classify 
relies on the posterior probability: 
$x$ is a sample $\sim$ 
$p_1$ rather than $\sim$ 
$p_0$ with probability
\begin{equation}\label{optimal_lambda}
    \mathrm{Pr}(k=1|x,\lambda, p_0, p_1) = \frac{\lambda p_1(x)}{\lambda p_1(x) + (1-\lambda)p_0(x)}.
\end{equation}
Indeed, as is well known 
(see e.g. \cite[Chap. 11]{giraud}), 
assigning a sample to the label with highest posterior probability is the optimal decision rule in the sense that it minimizes the probability of misclassification.

To compute this posterior probability, 
one needs to evaluate the 
pdfs
$p_1,p_0$ and 
know
the prior probability $\lambda$. 
Unfortunately, 
$\lambda$ is often unknown so \eqref{optimal_lambda} is intractable. If however we dispose of a set 
$\mathcal{D}=\{(x_i^{(k_i)}, k_i)\}_{i=1}^{N_0 + N_1}$ 
of labelled observations,
$\lambda$ can be estimated by $\frac{N_1}{N_1+N_0}$ where $N_1$ and $N_0$ are respectively the number of samples from $p_1$ and $p_0$. This leads to the (approximate) probability:
\begin{equation}\label{optimal_sans_lambda}
    \mathrm{Pr}(k=1|x, \mathcal{D}, p_0, p_1) = \frac{N_1p_1(x)}{N_1p_1(x) + N_0p_0(x)}. 
\end{equation}

However in most cases, 
$p_0$ and $p_1$ are unknown too
so \eqref{optimal_sans_lambda} cannot be computed either. 
When we only dispose of $\mathcal{D}$, 
we can make use of a parametric classifier
(in this paper we call 
classifier
any function $r_\phi(x)$ parameterized by $\phi$
which mimics the unknown posterior pdf).
% At this point, let us make clear what we mean by {\sl classifier};
% stricto senso,
% a classifier often denotes
% a measurable function on the observation space, with values in $\{0,1\}$
% \cite{giraud}.
% Of course, there exists classifiers which directly compute such a binary value function, such as the SVM method 
% \cite{foundations-ML}.
% However, as we just recalled,
% this binary function should ideally
% be the argmax of the posterior pdf.
% With a slight abuse of notation, 
% in this paper we will thus call 
% classifier, 
% any parameterized function $r_\phi(x)$
% which mimics the unknown posterior pdf.
%\eqref{optimal_lambda}.
So let us assume that 
%we can compute from $\mathcal{D}$ such a classifier, i.e.
we have at our disposal a function $r_{\phi}$ 
such that
\begin{equation}
\label{approx-classifier1}
    r_{\phi}(x) \approx \frac{N_1p_1(x)}{N_1p_1(x) + N_0p_0(x)}.
\end{equation}
Our paper is based on the observation that 
\eqref{approx-classifier1}
is equivalent to
\begin{equation}
\label{pdf_ratio_approx1}
    \frac{N_0}{N_1}\frac{r_{\phi}(x)}{1-r_{\phi}(x)} 
    \approx
    \frac{p_1(x)}{p_0(x)},
\end{equation}
which implies that (typically neural network based) classifiers can also be used for approximating 
{\sl pdf ratios}.

Equation 
\eqref{pdf_ratio_approx1}
has already been observed, 
and exploited in contexts where estimating a ratio 
of pdfs is relevant.
First, classifiers are at the core of adversarial training techniques in which divergence measures 
involving a ratio 
are replaced by an approximation based on a 
%classification model 
classifier \citep{nguyen2010estimating}. 
This enables to learn implicit generative models 
(i.e., with intractable pdfs) 
\citep{GAN} \cite{GANOverviewIEEE} \citep{creswell2018denoising}. 
Moreover, classifier based pdf ratio approximation has been applied to estimation of such metrics  as Mutual Information \citep{belghazi2018mutual}. 
Finally, 
classifiers based ratios have been applied successfully 
in statistical hypothesis testing procedures \citep{gretton2012kernel}, 
which heavily rely on likelihood-ratio tests.

If $p_0$ is an instrumental distribution with tractable pdf, 
then \eqref{pdf_ratio_approx1} can be turned into an approximation of target pdf $p_1$. 
So classifiers can be used for 
density estimation,
conditional density estimation, or likelihood-to-evidence ratio estimation, 
making them 
especially relevant 
in a likelihood-free inference setting \citep{durkan2020contrastive}\citep{hermans2020likelihood}\citep{thomas2022likelihood}.

However,
the question of sampling the corresponding model 
remains open,
and this is precisely the point we discuss in this paper. 
We realize that pdf ratios also play a key role in such simulation techniques as
the AR or 
Markov Chain Monte Carlo (MCMC) methods,
in which samples from instrumental $p_0$ are transformed into samples from the target $p_1$ via the ratio of the two densities.
This establishes a connection between
classification and MC sampling,
and will enable us to relax the assumption of tractable pdf $p_0,p_1$ of these sampling algorithms, 
at the price of approximate sampling. Our approach is therefore completely pdf-free, and as such is especially relevant when the target distribution is unknown or with intractable, noisy, or costly to evaluate pdf (see \citep{DBLP:journals/corr/abs-2108-00490} for a review of MC techniques in this setting, and \citep{sisson2018handbook} for a review of likelihood-free Approximate Bayesian Computation techniques); and/or when the instrumental $p_0$ is defined by a generative model with implicit pdf 
\cite{VAE}\citep{chen2018continuous}\citep{song2020score}\citep{ContinuouslyIndexedFlowGenerative}\citep{VIwithContinuouslyIndexedFlow}\citep{GAN}\citep{GANOverviewIEEE}.
The rest of this paper is organized as follows.
In \S
\ref{tutorial-simu}
we recall classical ratio-based stochastic simulation algorithms, i.e. 
the AR, IMH and IS techniques.
In \S \ref{standard_binary_classif}
we show that classifiers computed via the Binary Cross Entropy (BCE) criterion
indeed provide with an approximation of the posterior
\eqref{optimal_sans_lambda}.
Finally in \S \ref{connection simu-classifier}
we propose classification based sampling methods,
illustrate our method via simulations, and revisit it under the perspective of probabilistic modeling. 
We end the paper with a conclusion.

\section{Classical ratio-based 
sampling algorithms}
\label{tutorial-simu}

Stochastic simulation includes 
a variety of techniques,
see e.g.
\cite{bartoli-delmoral-2001}-\nocite{gentle-2004}\nocite{robert-casella}\nocite{Chib1995}\nocite{mcbook}\nocite{martino2018independent}\cite{hammersley-handscomb1964}.
In this section we focus on
AR, IMH and IS which share in common 
that they all compute a ratio of pdfs. 

\subsection{The AR algorithm}
\label{classical-AR}

\subsubsection{A brief reminder of AR Sampling}\label{rejection_sampling}

AR Sampling 
\cite[chap. 2]{robert-casella} \citep[chap. 3]{martino2018independent}
is a simulation algorithm 
that yields samples distributed according to a target distribution $p$ 
via samples from a proposal distribution $q$,
which are accepted or rejected as valid samples from $p$ 
via some acceptance probability.
More precisely, let the support of $p$ be inside that of $q$. 
This means that there exists a constant $C\geq1$ 
such that for all $y\in \mathbbm{R}^d, p(y) \leq Cq(y)$. 
Let $Y \sim q$, 
and let $k$ a Bernoulli random variable with parameter $\alpha_{AR}(y) = \frac{p(y)}{Cq(y)}$. 
AR sampling is based on the fact that $Y|k=1$ is distributed according to $p$.
Note that
$\mathrm{Pr}(k=1) = \frac{1}{C}$, 
so the lower the value of $C$, the higher the acceptance rate.

In order to use the algorithm in practice,
we thus need to know pdf $p$, 
and
build $q$ such that one can sample easily from $q$ and
there exists $C$ such that $p(y) \leq Cq(y)$ for all $y$, we can compute one such value of $C$,
and $C$ is as small as possible.
Note finally that
the algorithm can easily be adapted 
to the cases where $p$ and/or $q$
are known up to a (non necessarily common) constant,
see e.g. \cite[Th. 4.5]{mcbook}.

\subsubsection{Revisiting AR sampling 
as optimal binary classification}
\label{AR-classifier}

As we shall now see, 
AR sampling is indeed nothing but a binary classification procedure (see also \citep[\S 6]{castella:hal-00830124} for an application of this principle).
% in the context of a mixture model for blind source separation).
% Starting from the target pdf $p(x)$,
% we find a distribution $q$ 
% and constant $C>1$
% such that $C q(x)$ envelopes $p(x)$;
% % Since %the reminder 
% % $C q(x) - p(x)$ 
% % is non negative,
% % we write $C q(x)$ as $p(x)$ 
% % plus a positive reminder %$C q(x) - p(x)$
% % which,
% % up to a constant, is also a pdf;
% % so enveloping $p(x)$ with $C q(x)$
% % (or indeed $\frac{1}{C} p(x)$ with $q(x)$)
% by doing so we indeed build the implicit binary mixture distribution (see also figure \ref{fig:AR enveloppe} below) 
% %We start from the obvious eqality

Starting from the target pdf $p(x)$,
we find an easy-to-sample distribution $Q$ 
and constant $C>1$
s.t. $C q(x)$ envelopes $p(x)$.
Since %the reminder 
$C q(x) - p(x)$ 
is non negative,
we write $C q(x)$ as $p(x)$ 
plus a positive reminder %$C q(x) - p(x)$
which,
up to a constant, is also a pdf;
so enveloping $p(x)$ with $C q(x)$
% (or indeed $\frac{1}{C} p(x)$ with $q(x)$)
is nothing but building the implicit binary mixture pdf (see also figure \ref{fig:AR enveloppe} below) 
\begin{equation}
\label{AR-as-a-mixture}
\underbrace{q(x)}_{\rm proposal}= \frac{1}{C} \; \underbrace{p(x)}_{\rm target} 
\;+\; 
(1-\frac{1}{C})\;
\underbrace{
\frac{q(x)-\frac{1}{C}p(x)}{1-\frac{1}{C}}
}_{\rm reminder}\;
\end{equation}
with a priori probabilities $\frac{1}{C}$ and $1 - \frac{1}{C}$.
The first component of the mixture
is the target pdf $p$, 
and the second one is the law of the rejected samples.
\begin{figure}
    \centering
     \includegraphics[width=.9\linewidth]{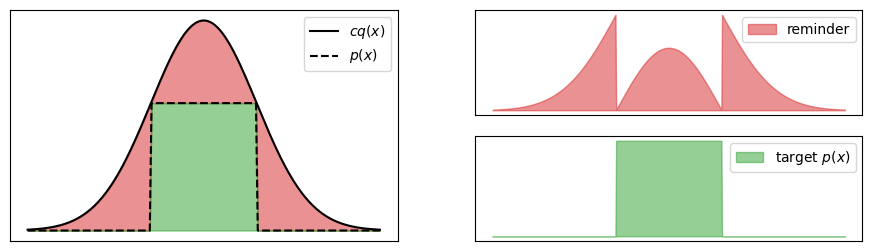}
    \caption{Envelopping target pdf builds an implicit mixture.}
    \label{fig:AR enveloppe}
\end{figure}
The magic of the AR algorithm consists 
in drawing samples 
from mixture  $q$
{\sl without} needing to sample from its two components
(see the r.h.s. of 
\eqref{AR-as-a-mixture}).
Accepting (or rejecting) a sample 
depending on the ratio probability 
\begin{equation}
\label{eqratio}
\alpha_{AR}(x) = \frac{p(x)}{C q(x)} =
\frac
{\frac{1}{C} p(x)}
{
\frac{1}{C} p(x)
+
(1-\frac{1}{C}) \frac{q(x)-\frac{1}{C}p(x)}{1-\frac{1}{C}}
}
\end{equation}
then amounts to classifying the samples 
with the posterior pdf
(compare 
\eqref{eqratio} to 
\eqref{optimal_lambda}).
\subsection{IMH}\label{IMH}

MCMC algorithms 
%is to build 
build a Markov chain whose invariant distribution is  
the target distribution $p$; 
so simulating the chain 
yields samples asymptotically distributed 
%according to 
$\sim$ $p$. 
The Metropolis-Hastings (MH) algorithm
\cite{robert-casella}
\cite{Chib1995}
is a particular MCMC method 
which constructs the Markov Chain $x_t$ as a two-step procedure: 
given a current state %of the chain 
$x_t$, the algorithm draws 
%proposes a new state 
a candidate $x^*$ from a proposal distribution $q(.|x_t)$, 
and then calculates the acceptance probability $\alpha_{MH}(x^{*}, x_t)=\min(1, \frac{p(x^*)q(x_t|x^*)}{p(x_t)q(x^*|x_t)})$. 
$x^*$ is accepted as the new state $x_{t+1}$ with probability $\alpha_{MH}(x^{*}, x_t)$; 
if $x^*$ is rejected then the chain remains in the current state $x_t$.
In practice, 
%the choice of Markov transition 
$q(.|x_t)$ plays a crucial role in the performance of the MH algorithm: 
if 
%this proposal distribution 
%$q(.|x_t)$ is 
not well-tuned, the acceptance rate may be too low, leading to slow mixing of the chain, or too high, leading to poor exploration of the target distribution.

The IMH algorithm is a simplified version of MH which considers an independent transition. The new point $x^*$ is hence proposed independently of the current state $x_t$, according to an independent proposal $q(.)$. 
In this case, the acceptance probability simplifies to $\alpha_{IMH}(x^{*}, x_t)=\min(1, \frac{p(x^*)q(x_t)}{p(x_t)q(x^*)})$.

\subsection{IS}
\label{reminder-is}

In many signal processing problems
we want to compute 
the expectation of some function $f$ with respect to pdf $p$: $\mu = \int f(x)p(x){\rm d}x = \E_{P}\bracket{f(x)}.$ In practice 
$\mu$
can be very difficult to compute,
so one needs to resort to approximations.
IS is a variance reduction technique
for integral MC estimates 
which can be traced back to the 1950's
\cite{Kahn1953}
\cite{marshall1956}
\cite[\S 5.4]{hammersley-handscomb1964}.

The crude MC estimate of $\mu$
reads
$\hat{\mu}^{\rm MC} = \frac{1}{N} \sum_{i=1}^N f(x_i)$ with 
$x_i \stackrel{\rm iid}{\sim} p$.
However it is generally difficult 
to sample directly from $p$,
moreover 
$\hat{\mu}^{\rm MC}$ can 
be a poor estimate,
particularly when the regions where $p$ is large 
do not coincide with those where $f$ is large.
Rewriting
$\mu = \int f(x) \frac{p(x)}{q(x)} q(x){\rm d}x$, where $q$ is some importance distribution, 
leads to the IS estimator $
\hat{\mu}^{\rm IS}(q) =
\frac{1}{N} \sum_{i=1}^N \frac{p(x_i)}{q(x_i)} f(x_i), \,\,\, 
x_i \stackrel{\rm iid.}{\sim} q.
$
As far as variance reduction is concerned,
one can easily show that
the importance pdf 
which minimizes
$\Var(\hat{\mu}^{\rm IS}(q))$
is 
$q_{\rm opt}^{\rm IS}(x) \propto |f(x)| p(x)$.
Even if in practice $\hat{\mu}^{\rm IS}(q_{\rm opt}^{\rm IS})$ cannot be computed,
this tells us that the regions where it is important to sample from
(whence the term "{\sl importance} distribution")
are not those where $p$ is large, 
but rather those
where $|f|p$ is large.
Note that 
$\hat{\mu}^{\rm IS}$
can be computed only if $p$ and $q$ are known exactly, or known up to a common constant; if this is not the case one can resort to self-normalized IS \cite{Geweke1989}.

Besides being a variance reduction technique, IS can also be seen as a two step sampling procedure for producing samples 
(approximatively) drawn from $p$,
out of samples originally drawn from $q$.
The technique is known as Rubin's
SIR mechanism
\cite{rubin1988},
\cite{gelfand-smith},
\cite{smith-gelfand},
\cite[\S 9.2]{Cappeetal}:
Let
$\{ x_i \}_{i=1}^N$ 
be $N$ iid. samples from $q(x)$,
and given $\{x_i\}_{i=1}^N$,
let $\{ \tilde{x}^i \}_{i=1}^M$ be $M$
iid. samples from 
$
\sum_{i=1}^N
\frac
{
%\frac{p(x_i)}{q(x_i)}
p(x_i)/q(x_i)
}
{
\sum_{i=1}^N 
%\frac{p(x_i)}{q(x_i)}
p(x_i)/q(x_i)
}
\delta_{x_i}({\rm d}x)$
(in other words, we draw samples from $q$,
weight each of them with weight proportional to $w^{u}(x_i) = p(x_i)/q(x_i)$,
and {\sl resample} $M$ iid. points from this random discrete probability mass function).
Then $\{ \tilde{x}^i \}_{i=1}^M$ are dependent and are not $p$-distributed,
but become iid. samples from 
$p({x})$
if $N \rightarrow \infty$.

\section{Parametric classifier by minimizing the BCE}
\label{standard_binary_classif}

From now on
we consider the setting where $\lambda$, $p_1$ and $p_0$
are  unknown,
and we only have the set 
$\mathcal{D}$ of labeled samples from
$p_0$ (with label $k=0$) and $p_1$ (with label $k=1$),
see \S \ref{intro}.
In this context, 
we should build a parametric function $r_{\phi}(x)$
that approximates the posterior pdf 
from the recorded samples. 
The aim of this section is to show that minimizing a BCE criterion indeed yields such a suitable approximation,
since the BCE, up to constants, is nothing but an MC approximation of a Kullback-Leibler Divergence ($D_{\mathrm{KL}}$) between the classifier and the unavailable posterior pdf.

To see this, let us first recall the BCE criterion:
\begin{equation}\label{BCE}
    \mathcal{L}_{\mathrm{BCE}}(\phi) = -\sum_{i=1}^{N_1} \log(r_\phi(x^{(1)}_i)) -  \sum_{i=1}^{N_0} \log(1 - r_\phi(x^{(0)}_i)),
\end{equation}
where $r_\phi(x) = \mathrm{Pr}(k=1|x,\phi)$ is the probability (under model $\phi$)
that the label associated to an observation $x$ is $1$. 

Let $h(x,k)$ be the joint distribution over observations and labels: 
\begin{align}
    \label{joint}
    &h(x,k)  \!=\! \underbrace{\frac{N_k}{N_1+N_0}}_{h(k)}\underbrace{p_k(x)}_{h(x|k)}, x\in\mathbbm{R}^d, k=0,1.
\end{align}
On the other hand, using $r_\phi(x)$, 
we construct another joint probability distribution 
$
h_\phi(x,k) = h(x) r_{\phi}(x)^{k}(1-r_{\phi}(x))^{1-k}
$,
 where $h(x)$ is the $x$-marginal in \eqref{joint}. 
 As the name Cross-Entropy suggests, the BCE loss is, up to additive and multiplicative constants, nothing but an MC approximation of $D_\mathrm{KL}\paren{h(x,k)||h_\phi(x,k)} = \E_{h(x)}\paren{D_\mathrm{KL}(h(k|x)||r_{\phi}(x)^{k}(1-r_{\phi}(x))^{1-k})}$, 
see appendix.

The interest of this interpretation 
is that,
%we can easily deduce the expression of the optimal classifier
%(that for which the $D_\mathrm{KL}$ reaches its minimum)
as is well known, 
a $D_{\mathrm{KL}}$ reaches zero if and only if the two distributions are equal almost surely.
So, if $r_\phi$ represented any arbitrary function, minimizing 
$D_\mathrm{KL}\paren{h(x,k)||h_\phi(x,k)}$
would ensure that $r_{\phi}(x)^{k}(1-r_{\phi}(x))^{1-k} = h(k|x)$ for all $x\in\mathbbm{R}^d$ and for $k=1,0$,
i.e. that the classifier reaches the target posterior pdf. Of course, in practice, 
minimizing the BCE does not ensure that
this  $D_\mathrm{KL}$ decreases to zero. 
First, since we only dispose of a finite number of labeled observations, minimizing an MC approximation of the $D_{\mathrm{KL}}$ does not minimize the $D_{\mathrm{KL}}$ itself.
% guarantee that the $D_{\mathrm{KL}}$ itself is minimized.
Next, the parametric family does not contain $h(k|x)$ in general, 
in which case we can only ever reach a positive minimum of the $D_{\mathrm{KL}}$. 
Lastly, standard optimization techniques would only guarantee convergence to a local minimum of the $D_{\mathrm{KL}}$. 
Therefore in practice,
minimizing the BCE loss only provides with $r_{\phi}$ which approximates the unknown posterior.

\section{Using a binary classifier for (approximate) Sampling}\label{parametric_approx}
\label{connection simu-classifier}

We now come to the heart of this paper.
If $p_1$ is a pdf of interest in 
an MC sampling setting, and $p_0$ a suitable 
easy-to-sample instrumental distribution
- be it the proposal distribution in AR, the independent Markov transition Kernel in IMH, or the importance distribution in IS, then the three sampling algorithms involve the pdf ratio $p_1(x)/p_0(x)$, which is unknown when at least one pdf is intractable.
As explained in section
\ref{standard_binary_classif},
a parametric binary classifier
trained from a set $\mathcal{D}$ of labeled observations
computes an approximation 
of the unknown posterior distribution.
However,
remember that 
\eqref{approx-classifier1}
is equivalent to
\eqref{pdf_ratio_approx1};
we thus see that classifiers can also be used for approximating {\sl pdf ratios} of interest,
which enables us to propose
approximate versions of the sampling algorithms based on this classifier-ratio approximation,
and thus to relax the requirement of tractable pdf, but at the cost of approximate sampling.
Of course, the closer $p_0$ is to $p_1$, the more efficient the algorithms. However, here $p_0$ is supposed to be given and hence, our problem is not (as usual) to adjust $p_0$ from a given $p_1$,
but to make the most of $\mathcal{D}$ 
for fixed $p_0, p_1$.

\begin{flushleft}
{\sl Assumptions.}
\end{flushleft}
$p_1$ is the distribution of interest and $p_0$ a fixed instrumental distribution from which we can propose samples.
Ratio $p_1(x)/p_0(x)$ is unknown and we dispose of the labeled dataset $\mathcal{D}$, and assume that we can train a binary classification model $r_{\phi}$ which minimizes \eqref{BCE}.

\subsection{Classifier-based sampling algorithms}
Remember that a  key ingredient 
for running the algorithms 
of \S \ref{tutorial-simu}, 
% with target $p_1$ and instrumental $p_0$, 
is the ratio $p_1(x)/p_0(x)$, 
which appears in $\alpha_{AR}(x), \alpha_{IMH}(x, x_t)$ and in $w^u(x)$.
% But in our setting, 
% this ratio cannot be computed 
% since $p_1(x)$ and/or $p_0(x)$ is unavailable.
Following the idea expressed 
in 
\eqref{pdf_ratio_approx1},
we can however make use of a classifier 
for approximating the unavailable ratio
$p_1(x)/p_0(x)$, and finally the quantities: 
\begin{align}
    &\alpha_{AR}(x) \leftarrow 
    \frac{1}{\widetilde{C}} 
    \frac{r_{\phi}(x)}{1-r_{\phi}(x)} \text{ where }
    \widetilde{C} = \max_{y\in \mathcal{D}}
    \frac{r_{\phi}(y)}{1-r_{\phi}(y)}; \label{ar_approx}
    \\
    &\alpha_{IMH}(x, x_t) \leftarrow  \min\paren{1, \frac{r_\phi(x)(1-r_\phi(x_t))}{(1-r_\phi(x))r_\phi(x_t)}}; \label{imh_approx}\\
    & w^{\rm u}(x)
    \leftarrow
    \frac
    {r_\phi(x)}{1-r_\phi(x)}.\label{is_approx}
\end{align} 
\begin{figure}
    \centering
     \includegraphics[width=\linewidth]{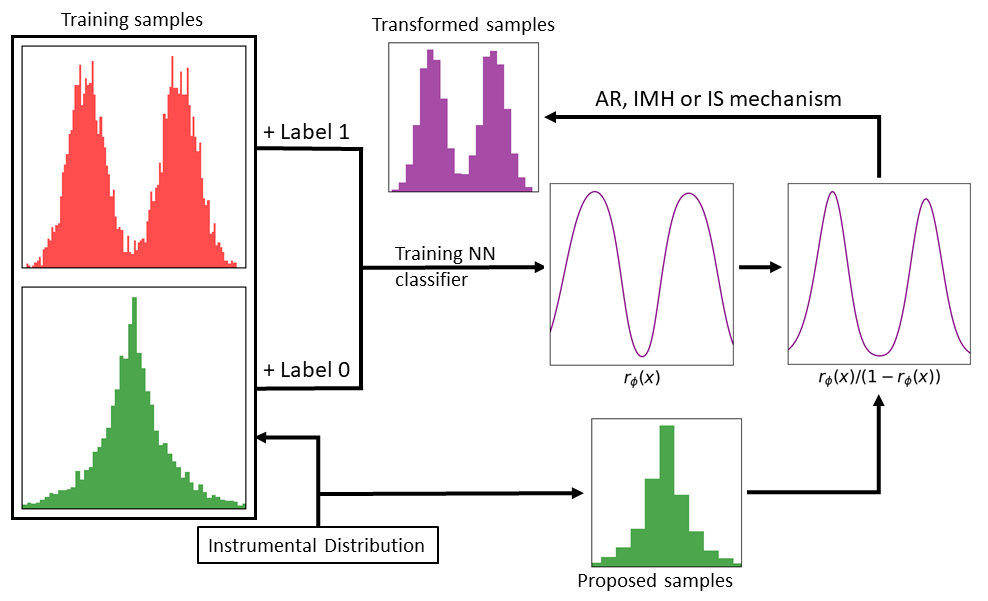}
    \caption{Summary of the classifier based sampling approach}
    \label{fig:summary}
\end{figure}
Our procedure is summarized by Figure \ref{fig:summary}:
we first train $r_\phi$
from labeled samples from $p_1$ and $p_0$;
we next use ratio $r_\phi(x)/(1-r_\phi(x))$
as a surrogate of 
$p_1(x)/p_0(x)$,
which enables us to use 
the AR, IMH or IS procedure,
and thus to turn
samples from $p_0$ 
into (approximate) samples
from $p_1$.
A main advantage of our approach is that a distribution which is only defined by its sampling procedure and has implicit intractable pdf can be used as instrumental $p_0$. Indeed our approach does not require evaluating the pdf $p_0$ neither during the training of the classifier, nor during the three proposed sampling procedures.

\subsection{Illustrating examples} 
%\subsection{Illustration on Toy Examples}
\begin{figure}
    \centering
    \begin{subfigure}[b]{0.16\linewidth}
        \includegraphics[width=\linewidth]{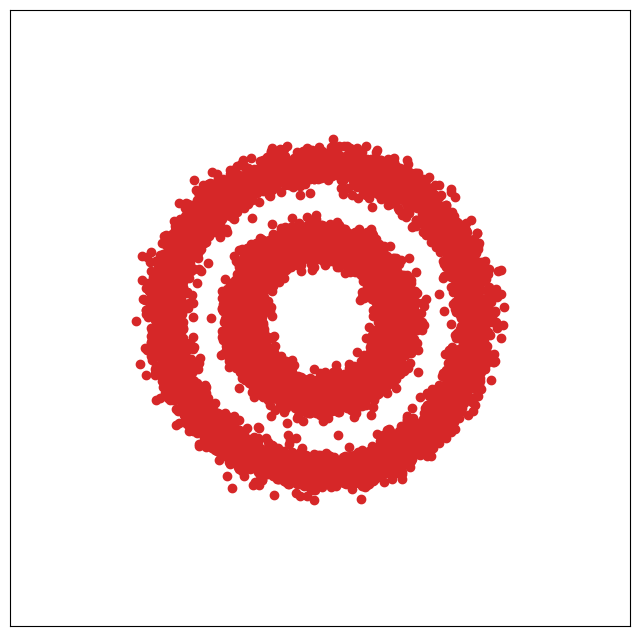}
    \end{subfigure}
    \begin{subfigure}[b]{0.16\linewidth}
        \includegraphics[width=\linewidth]{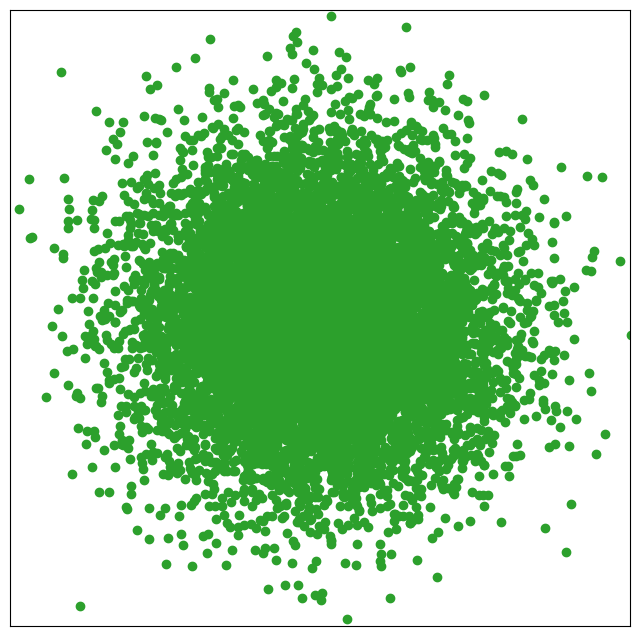}
    \end{subfigure}
    \begin{subfigure}[b]{0.16\linewidth}
        \includegraphics[width=\linewidth]{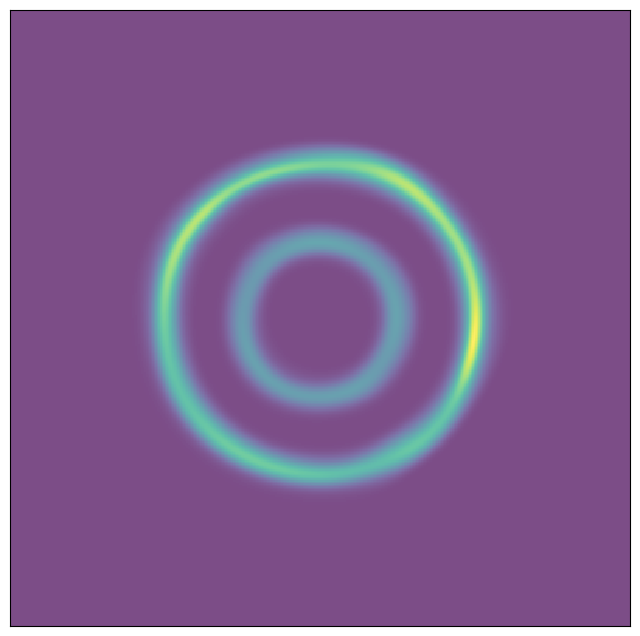}
    \end{subfigure}
    \begin{subfigure}[b]{0.16\linewidth}
        \includegraphics[width=\linewidth]{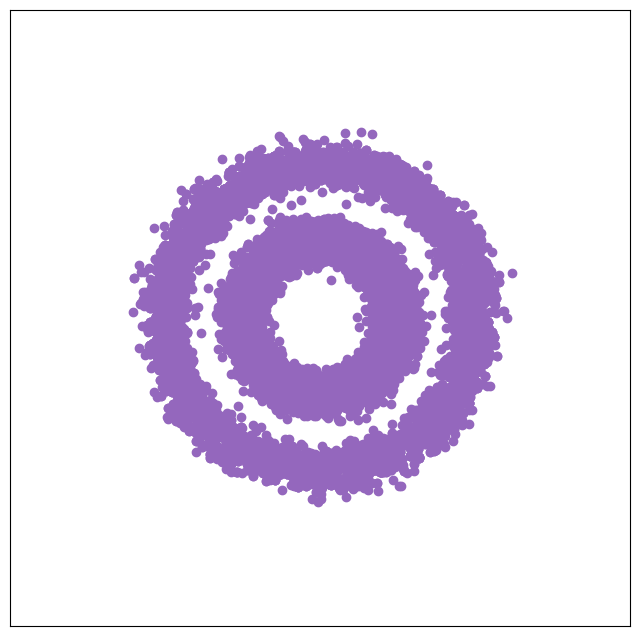}
    \end{subfigure}

        \centering
    \begin{subfigure}[b]{0.16\linewidth}
        \includegraphics[width=\linewidth]{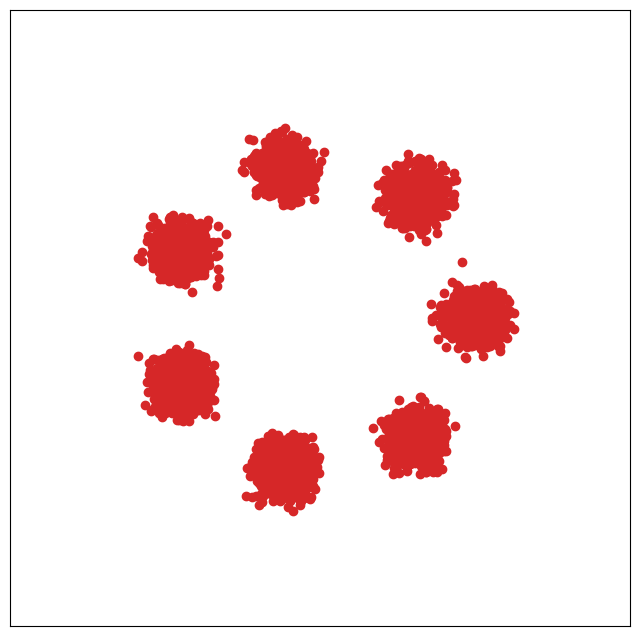}
    \end{subfigure}
    \begin{subfigure}[b]{0.16\linewidth}
        \includegraphics[width=\linewidth]{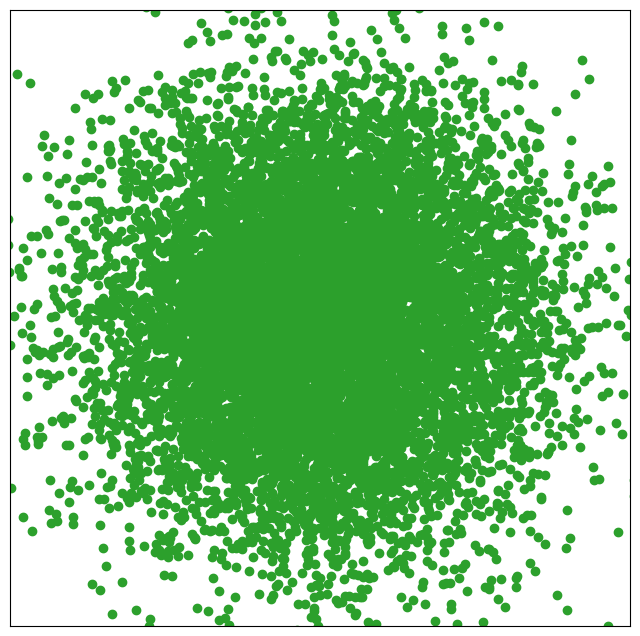}
    \end{subfigure}
    \begin{subfigure}[b]{0.16\linewidth}
        \includegraphics[width=\linewidth]{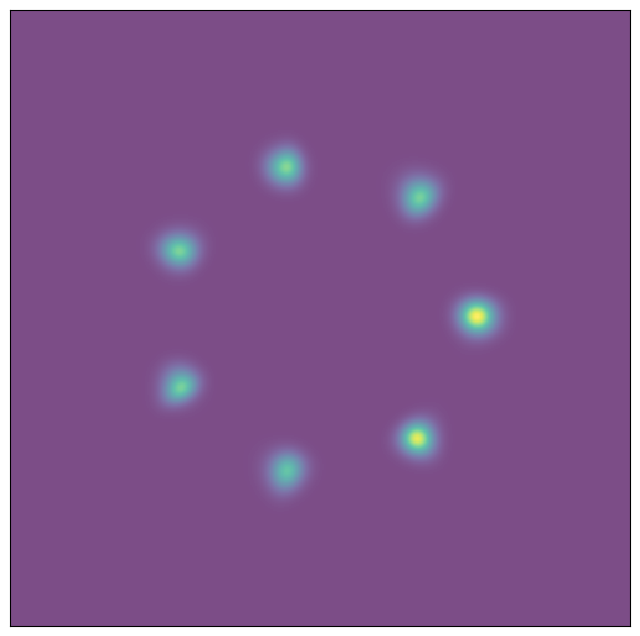}
    \end{subfigure}
    \begin{subfigure}[b]{0.16\linewidth}
        \includegraphics[width=\linewidth]{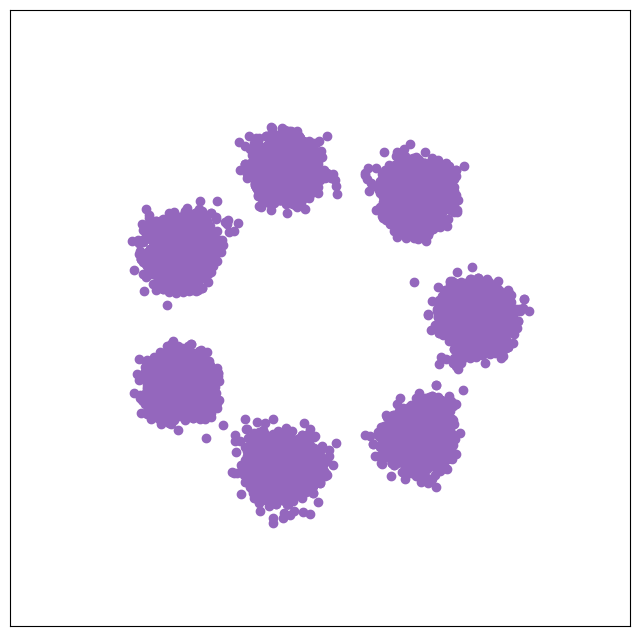}
    \end{subfigure}

        \centering
    \begin{subfigure}[b]{0.16\linewidth}
        \includegraphics[width=\linewidth]{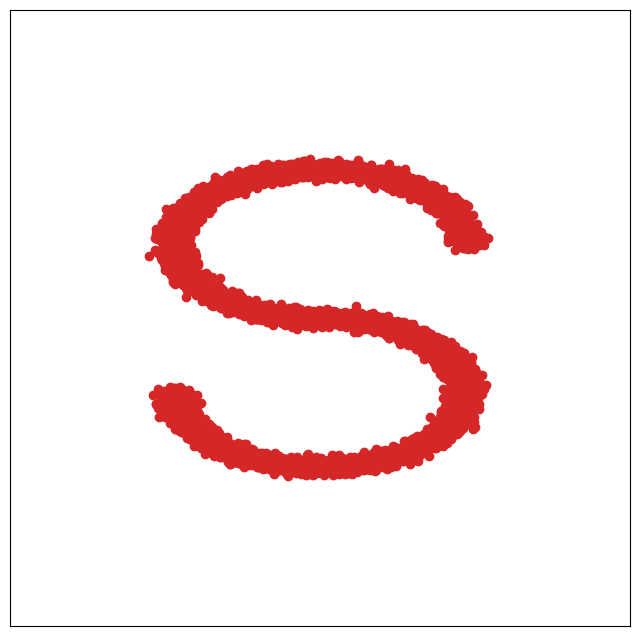}
    \end{subfigure}
    \begin{subfigure}[b]{0.16\linewidth}
        \includegraphics[width=\linewidth]{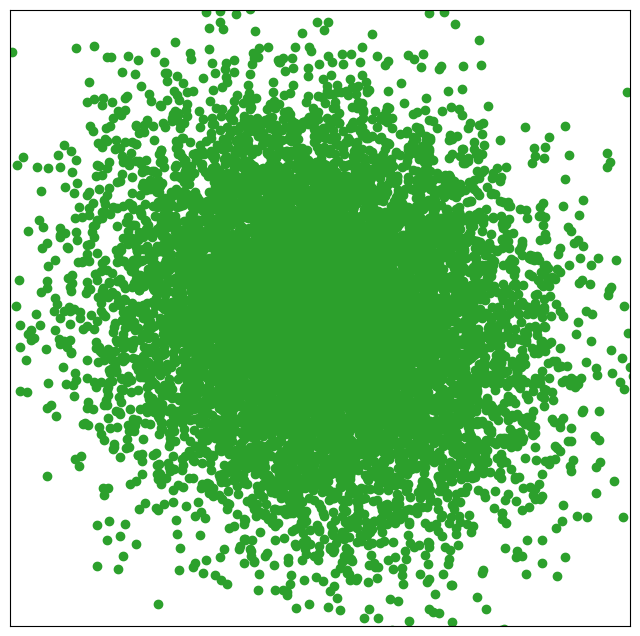}
    \end{subfigure}
    \begin{subfigure}[b]{0.16\linewidth}
        \includegraphics[width=\linewidth]{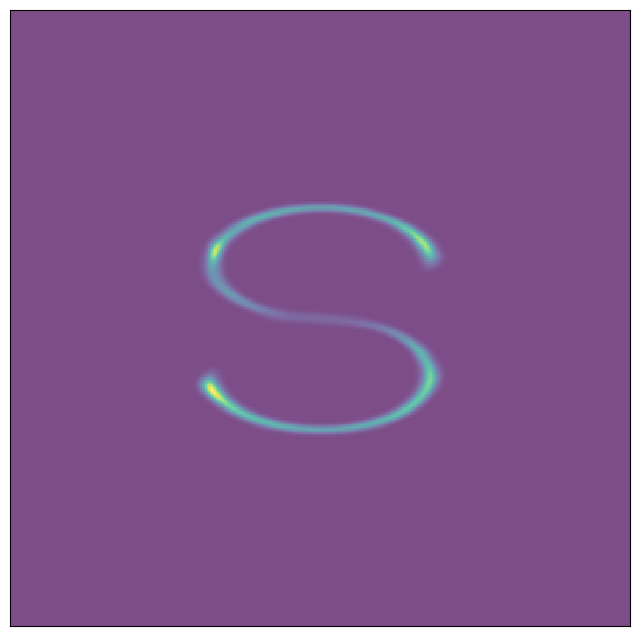}
    \end{subfigure}
    \begin{subfigure}[b]{0.16\linewidth}
        \includegraphics[width=\linewidth]{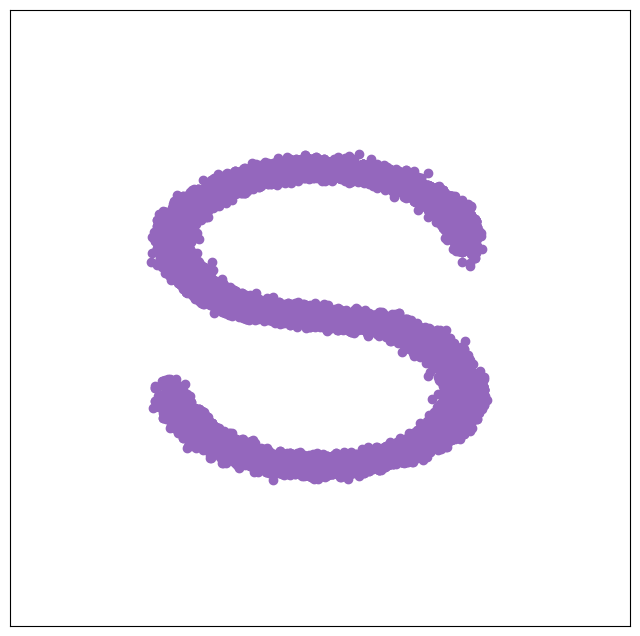}
    \end{subfigure}
    \caption{Density ratio (middle-right) via classification of samples from $p_1$ (left) and $p_0$ (middle-left) - approximate samples from $p_1$ (right) obtained via a ratio based algorithm: AR (top), IMH (Middle), IS (bottom)}
    \label{toy_example}
    \label{illustration}
\end{figure}
We illustrate our approach (see fig. \ref{illustration}) on reference 2D examples in order to illustrate the mechanism of (i) obtaining an approximate of the pdf ratio from samples 
using a feed-forward neural network \citep{bebis1994feed} with 3 hidden layers, 32 hidden units per layers and SiLU activation function that outputs $\mathrm{logit}(r_\phi(x))$; and (ii) sampling from the target distribution via that pdf ratio using the AR, IMH or IS samplers.
The instrumental $p_0$ was set to be Gaussian with mean and covariance estimated from the samples from $p_1$ (even though it can be computed, pdf $p_0$ was not used during the procedure).

\subsection{Probabilistic modelling}
So far, we have presented our work as a technique to perform approximate MC sampling; 
let us now revisit 
it under the scope of probabilistic modelling. If we rewrite $p_1$ as
\begin{equation}
    p_1(x) = \frac{p_0(x)(p_1(x)/p_0(x))}{\int p_0(z)(p_1(z)/p_0(z))\mathrm{d}z},
\end{equation}
then using \eqref{pdf_ratio_approx1} amounts to building an approximation $p_\phi$ of $p_1$:
\begin{equation}\label{p_phi}
    p_\phi(x) = \frac{p_0(x)(r_{\phi}(x)/(1-r_{\phi}(x)))}{\int p_0(z)(r_{\phi}(z)/(1-r_{\phi}(z)))\mathrm{d}z}.
\end{equation}
Our procedure consists in applying the AR, IMH or IS samplers to $p_\phi$ with proposal $p_0$ (at least up to the approximation of constant $C$ in the AR case).
This construction corresponds to a specific energy-based model \citep{zhai2016deep}\citep{carreira2005contrastive}\citep{hinton2012practical}
with energy function $E_\phi(x) = - \log(p_0(x)) - \rm{logit}(r_\phi(x))$. 
Model $p_\phi$ inherits the advantages of this energy structure:
(i) it can be trained without evaluating the gradient of the 
numerator of \eqref{p_phi} nor of the intractable normalizing constant; (ii) it is structurally compatible with the  AR, IMH or IS samplers with proposal $p_0$.

\section{Conclusion}
In this paper 
we proposed a version of the classical AR, IMH or IS samplers,
with target $p_1$
and proposal $p_0$,
in which 
the key $(p_1/p_0)$ ratio is replaced by a surrogate function trained from a labelled dataset.
From an MC perspective, the advantages or our approach are threefold:
(i) it is completely pdf-free;
(ii) training amounts to building a
(typically neural network based) classifier;
(iii) the instrumental 
pdf $p_0$ does not need to be known explicitely.
From a probabilistic modeling perspective,
our approximate samplers coincide with the original ones when applied to some specific energy based 
approximation
of target $p_1$ which,
thanks to its specific structure,
can both be trained easily via standard classification, 
and is structurally compatible with the AR, IMH or IS sampling techniques.

\appendix
\vspace{-15pt}
\begin{eqnarray}
\nonumber
D_\mathrm{KL}(h(x,k)||h_\phi(x,k))
=
\E_{h(x,k)}[\log(h(x,k))] \\
-\E_{h(\!k,x)}[\log\paren{h(x)}] 
-\E_{h(k,x)}[\log(\mathrm{Pr}(k|x,\phi))].\label{lastterm}
\end{eqnarray}
In \eqref{lastterm}, only the last term depends on
$\phi$; we get the BCE loss with an MC approximation of it
(or, equivalently, 
replacing the expectation with one computed on the empirical distributions):
\vspace{-5pt}
\begin{align*}
    &\!\E_{h(k,x)}[\log(\mathrm{Pr}(k|x,\phi))] \!\stackrel{\eqref{joint}}{=} 
    \!\sum_{k = 0}^1\!\!\frac{N_k}{N_1\!\!+\!\!N_0}\!\!\int\!\log(\mathrm{Pr}(k|x,\phi))p_k(x) \mathrm{d}x \\
    \approx & 
    \sum_{k = 0}^1\frac{1}{N_1\!\!+\!\!N_0}  \sum_{i=1}^{N_k}
    \log(\mathrm{Pr}_\phi(k|x_i^{(k)}))\\
    = & 
    \frac{1}{N_1\!\!+\!\!N_0}
    \underbrace{(\sum_{i=1}^{N_1}\log(r_\phi(x^{(1)}_i)) +  \sum_{i=1}^{N_0}\log(1- r_\phi(x^{(0)}_i)))}_{-\mathcal{L}_{\mathrm{BCE}}(\phi)}.
\end{align*}
So
$
D_\mathrm{KL}\paren{h(x,k)||h_\phi(x,k)}\approx
A +
B\mathcal{L}_{\mathrm{BCE}}(\phi),
$
and
$
\arg\min_{\phi} D_\mathrm{KL}\paren{h(x,k)||h_\phi(x,k)} 
$
$\approx$
$\arg\min_{\phi}\mathcal{L}_{\mathrm{BCE}}(\phi).
$

\bibliographystyle{ieeetr}
\bibliography{bibliography}

\end{document}